\documentclass[a4paper,11pt]{article}
\pdfoutput=1 

\usepackage{jcappub} 

\usepackage[T1]{fontenc} 
\usepackage{gensymb}
\usepackage{verbatim}

\title{\boldmath Searching for Extremal Spots in \textit{Planck} Lensing Maps}

\author[a,b,1]{Clemens Jakubec,\note{Corresponding author.}}
\author[b]{Raelyn M. Sullivan,}
\author[b]{Douglas Scott}

\affiliation[a]{Department of Physics \\Imperial College\\ Prince Consort Road, London SW7 2AZ, UK}
\affiliation[b]{Department of Physics \& Astronomy \\ University of British Columbia \\ Vancouver, BC V6T 1Z1, Canada}

\emailAdd{clemens.jakubec16@ic.ac.uk}
\emailAdd{rsullivan@phas.ubc.ca}
\emailAdd{dscott@phas.ubc.ca}

\abstract{A great deal of attention has been given to the so-called Cold Spot in maps of the cosmic microwave background (CMB) temperature.  
We present a similar analysis, searching for extremal spots in the CMB lensing convergence and lensing potential maps from the {\it Planck\/} 2018 data release.
We perform a multi-scale and multi-filter analysis using the first three members of the Mexican-hat wavelet family to search for extremal features of different shapes and sizes.  Although an initial analysis appears to show the existence of some extremal spots at scales below about $5\degree$, we conclude, after marginalising over all scales and filters, that no significant features are detected in the lensing maps. We conclude that in terms of maxima and minima of various sizes, the lensing data have similar statistical properties to Gaussian simulations.}

\begin{document}
\maketitle
\flushbottom

\section{Introduction}
\label{Introduction}
 Features in the cosmic microwave background (CMB) at large angular scales,
 often referred to as `anomalies', continue to be a focus of attention in cosmology. This is because of the enticing possibility that one of them might be evidence for new physics on the very largest scales.
 Perhaps the most extensively studied of these large-scale CMB features is the so-called `Temperature Cold Spot', which we abbreviate as `TCS'.  It was first discovered \cite{cold_spot_1} in temperature anisotropy data from the Wilkinson Microwave Anisotropy Probe ({\it WMAP\/}) \cite{bennett2010}, manifesting as a cool region that appears deeper than expected at a scale of around $5\degree$. This feature was confirmed in data from the \textit{Planck} satellite mission \cite{Planck_13_Isotropy, Planck_15_Isotropy, Planck_18_Isotropy}, which had higher sensitivity and better removal of foreground emission sources (due to its broad frequency coverage).  There can therefore be little doubt that the TCS is a genuine feature on the CMB sky.  However, there remains disagreement as to whether or not the structure is statistically unusual (see for example refs.~\cite{hansen2007,schwarz2016,frolop2016}).  One way to address this is to search for related features in independent data sets.
 
 CMB polarization data represent one avenue of investigation that has been explored in ref.~\cite{Planck_18_Isotropy}; however, in practice the \textit{Planck} polarization data are not sensitive enough to provide an independent test of the existence of a specific feature on a scale of $5\degree$.  Nevertheless, there is one more data product derived from the CMB that can be examined, namely maps of the lensing potential or convergence. Such maps have been available since 2013
 \cite{Planck_13_Lensing}, and it is only natural to ask whether anything like the TCS exists in these lensing data.  Although this may seem premature, since these maps are still fairly noisy, in fact the total signal-to-noise ratio in the \textit{Planck} lensing maps is already superior to that of the first CMB temperature map from the \textit{COBE} satellite, which was subject to a battery of tests of statistical
 isotropy and Gaussianity \cite{bromley1999}.  The quality of lensing data is expected to improve considerably with the next generation of ground-based experiments \cite{SO,CMBS4} and and hence future analyses will be able to carry out even more sensitive tests.

 The \textit{Planck} lensing maps have been used to search for
 correlations with other data sets \cite{hill2014,Planck_CIB_lensing,bianchini2015,kirk2016,cai2017,singh2017,han2019}, but few analyses have been carried out on the lensing data themselves.
 Some exceptions include the analysis of the 1-point distribution of the maps and tests of the isotopy of the 2-point function \cite{Planck_13_Lensing}, as well as the variance \cite{marques2018}.  In this paper we focus explicitly on searching for TCS-like features in the lensing data.
 
 There are two versions of the lensing products available, namely the convergence and the potential, either of which could be argued to be the more physical quantity to study.  The spherical harmonic transforms of these maps simply differ by two powers of multipole, $\ell$, and hence they are essentially two representations of the same data, differing only in the weighting of small scales relative to large scales.  Obviously one could also construct further maps differing by other powers of $\ell$.
 Here we choose to mainly consider the\textit{Planck} 2018 lensing-convergence map, first looking explicitly in the direction of the TCS and then performing a wider search for extremal features. We specifically carry out a multi-scale analysis using the first three members of the Mexican-hat wavelet family~\cite{Planck_18_Lensing} over a range of angular scales.  In appendix~\ref{Lensing Potential} we provide the results of the same analysis performed on the lensing-potential map rather than the convergence map.
 
This paper is arranged as follows.
In section~\ref{TheCMBcoldspot}, we give some background information on the temperature Cold Spot, followed, in section~\ref{Method}, by a description of the filters used in this paper to search for spots in the lensing maps. In section~\ref{Mexhat} we introduce the methods and statistical approach used to look for extrema in the filtered lensing maps, as well as describing how we mask the data. In section~\ref{Results}, we present the results of an (unmarginalised) analysis of the lensing convergence, followed by our final results of an analysis that includes marginalisation over all filters and scales.  We conclude in section~\ref{Conclusion}.

\section{The Temperature Cold Spot}
\label{TheCMBcoldspot}

\begin{figure}[h!]
\centering
\includegraphics[width=0.49\linewidth]{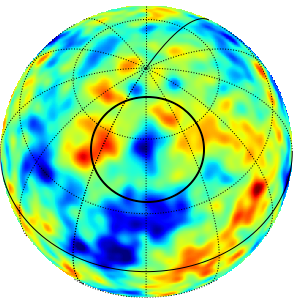}
\hfill
\includegraphics[width=0.49\linewidth]{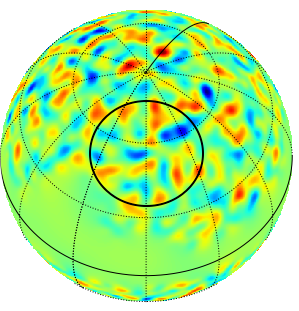}
\caption{\label{cold_spot}Left: CMB temperature map (explicitly from the \texttt{Commander} component-separation method) filtered with a Gaussian having a standard deviation of $2\degree$ and rotated such that the Cold Spot at Galactic coordinates $(l,b) = (210\degree, -57\degree)$ is at the centre of the map. The black circle lies just outside the positive rim of the Cold Spot. Right: CMB lensing convergence (also filtered with a $2 \degree$ Gaussian). No statistically significant Cold Spot-like feature can be seen near the centre of this lensing map.}
\end{figure}

As mentioned in the introduction, the TCS was first discovered in data from the \textit{WMAP} experiment and confirmed in each data release from the Planck Collaboration.  A detailed empirical description of the TCS can be found in refs.~\cite{vielva2010,cruz2010}.  It appears as a cool region, perhaps 5--$10\degree$ across, surrounded by a hotter ring, giving the appearance of a fairly circular `compensated' perturbation on the CMB sky (see the left panel of figure~\ref{cold_spot}).  The TCS is not actually the coldest place on the CMB sky, and it does not stand out particularly strongly when using a purely Gaussian-shaped filter. The feature is most conspicuous when using a matched filter of the `Mexican-hat' type with a scale of around $5\degree$.  In some studies it has been argued that this filter shape is a natural or ideal one to select, and it has been claimed that even when searching over a range of filter scales the likelihood of finding such a feature in random CMB skies is small (e.g.~refs~\cite{cruz2005,cruz2007a}). On the other hand, other papers have pointed out that when allowing for the possibility of having features with a wider range of shapes, the significance becomes much less remarkable (e.g.~ref.~\cite{zhang2010}).

Despite this debate regarding the significance of the TCS, there have been many suggestions of physical explanations for such a feature on the sky.  Proposed explanations include a giant void \cite{inoue2006,rudnick2007,szapudi2014,zibin2014,kovacs2016,mackenzie2017}, cosmic bubble collisions \cite{chang2009}, a cosmological texture \cite{cruz2007b}, a cosmic string loop \cite{ringeval2016} and modified inflationary scenarios \cite{afshordi2011}.  For some of these explanations there have been calculations of the predicted signals in polarization or lensing data \cite{das2009,czech2010,vielva2011}.

\begin{figure}[h!]
\centering 
\includegraphics[width=.50\textwidth]{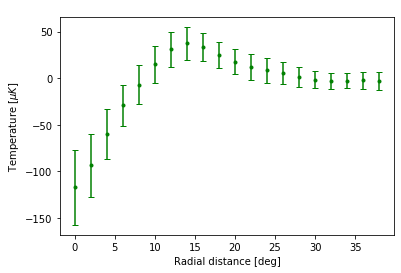}
\hfill
\includegraphics[width=.49\textwidth]{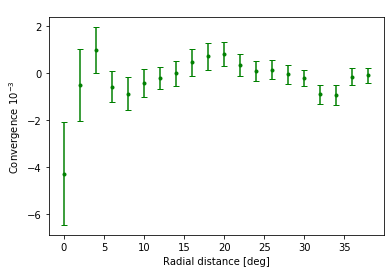}
\caption{\label{cold_spot_profile}Left: Temperature profile of the TCS. Right: Profile of the \textit{Planck} lensing map around the same point in the sky where the Cold Spot is found in the temperature map.  The error bars here are the dispersions from 50 annuli around random positions.}
\end{figure}

Given all the attention devoted to this particular sky direction,
an obvious question to ask is if the lensing maps exhibit a similar structure to the TCS in the same place as the temperature maps. Figure~\ref{cold_spot} shows the temperature (left) and lensing convergence (right), with the centre of the TCS at Galactic coordinates $(l,b) = (210\degree, -57\degree)$ rotated to the centre of the plot.  With some smoothing to suppress smaller-scale structure, the TCS can easily be identified in the temperature map; however, no particularly unusual feature can be seen in the lensing map. Figure~\ref{cold_spot_profile} in turn shows the profile of the temperature map (left) and the lensing-convergence map (right) around this same position, which was found by binning the data in 20 evenly spaced annuli between $0\degree$ and $40\degree$ around the centre of the TCS. The errors were estimated by choosing 50 random points on the sky, calculating the profile around those points like for the TCS and then finding the dispersion of the 50 averages for each annulus. Taking into account the error bars, the TCS exhibits a shape similar to that of a Mexican-hat filter, whereas the lensing convergence does not. The slightly low convergence around $(210\degree, -57\degree)$ is due to a slightly negative area in the lensing map, of no significance and there is clearly nothing of the size and shape of the TCS in this direction in the {\it Planck\/} lensing map.

An additional test we can undertake is to ask whether there are any significant features in the lensing map that are similar in shape to the TCS itself. To do this we use a circular model following the profile in figure~\ref{cold_spot_profile}. When we filter the sky with this shape we find that the position with the most extreme peak lies at $(l,b)=(245\degree,41\degree)$, with a value of $3.3$ times the standard deviation of the entire map.  However, it is not statistically unlikely to find such an amplitude, since 87\,\% of simulations exhibit a deviation larger than $3.3\,\sigma$ somewhere on the sky.

In the rest of this paper we carry out a general search for features of a range of shapes and sizes in the \textit{Planck} lensing map.
Previous investigations have focused on looking for cold spots in CMB temperature maps using Mexican-hat shaped filters \cite{vielva2010}, which works well for detecting the TCS because of its compensated shape. In accordance with those studies we also choose members of the Mexican-hat wavelet family to study the lensing data. It is important to realise that there is no physical model that predicts that interesting features are explicitly the shape of a Mexican-hat wavelet; however, they provide a useful basis set for searching for extremal spots of a range of shapes. It is not a theoretically predicted shape but it is as good a test as any other for detecting unusual features on the sky.  There is also no firm rules for deciding how many different filter shapes to use in any search.  We find that testing for three families over a wide range of angular scales is sufficient for our purposes, although of course one could always add more distinct filter shapes.

\section{Mexican-hat wavelets}
\label{Mexhat}

\begin{figure}[h!]
\centering 
\includegraphics[width=.5\textwidth]{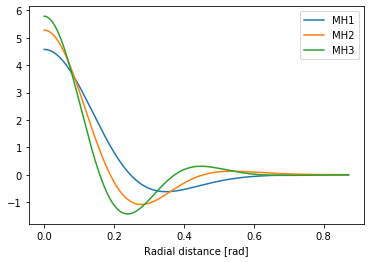}

\caption{\label{3MHWS} The first three members of the Mexican-hat wavelet family, generated as explained in section~\ref{Mexhat}}
\end{figure}

In order to study structures in the lensing map at different scales, it is necessary to apply a set of filters to the data. The filters of choice in this paper are the first three members of the Mexican-hat wavelet (MHW) family, which have been used extensively to study features in the CMB ~\cite{Mexhat_1}. The MHW family is defined in real space as 

\begin{equation}
\label{MHWF}
\psi_n(x)\propto \nabla^{2n} \phi(x),
\end{equation}

\noindent where in our case $\phi$ is a 2-d Gaussian, $\phi(x)=\exp{(-x^2/2)}/2\pi$, with unit standard deviation and $\nabla^{2n}$ is the Laplacian applied $n$ times. The reason the MHW set is usually chosen is because it satisfies the compensation condition 

\begin{equation}
    \label{compensation}
    \iint dx^2 \psi_n(x)=0,
\end{equation}

\noindent where the the double integral is over all 2-d space. Physically this means that as the map is convolved with the filter, the total area sums to zero, with the negative rim compensating for the positive peak. Applying the above definition, the three wavelets used in this paper are explicitly given by:
\begin{subequations}\label{MHW}
\begin{align}
\label{MHW1}
\psi_1(x)&=\frac{1}{\sqrt{2\pi}R} \left(2-\frac{x^2}{R^2}\right) \exp\left(-\frac{x^2}{2R^2}\right),
\\
\label{MHW2}
\psi_2(x)&=\frac{1}{\sqrt{24\pi}R}\left(8-\frac{8x^2}{R^2}+\frac{x^4}{R^4}\right) \exp\left(-\frac{x^2}{2R^2}\right),
\\
\label{MHW3}
\psi_3(x)&=\frac{1}{\sqrt{720\pi}R}\left(-48+72\frac{x^2}{R^2}-18\frac{x^4}{R^4}+\frac{x^6}{R^6}\right)\exp\left(-\frac{x^2}{2R^2}\right).
\end{align}
\end{subequations}
The coefficients are chosen such that $\iint d x^2 |\psi_n(x)|^2=1$. Here $R$ represents the scale at which the sky is filtered. The MHW effectively compares the value of the map over some central region with a local average in the rim of the hat. All three wavelets are shown in figure~\ref{3MHWS}. Since we analyse the sky at relatively small scales of less than $20\degree$, we work in the flat-sky approximation, although spherical Mexican-hat wavelets have been studied and used before~\cite{Mexhat_2}. 

\section{Methods}
\label{Method}
\subsection{Data and analysis}

The analysis presented here is performed on \texttt{HEALPix}\footnote{\url{http://healpix.sourceforge.net}} maps with $N_{\text{side}}=128$ \citep{gorski2005}. We use the \textit{Planck} 2018 CMB lensing-convergence map \cite{Planck_18_Lensing} built from the minimum-variance quadratic estimator (and the cosmic infrared background component-separated map to provide additional information on small scales) and the associated simulation package containing 300 simulations available from the Planck Legacy Archive.\footnote{\url{http://pla.esac.esa.int/pla}}  The spherical harmonic coefficients are provided for both the data and the simulations up to lensing multipole $L=2000$, with $L<9$ modes removed. The mask is also provided at $N_\text{side}=2048$ resolution. 

\begin{figure}[htbp!]
\centering 
\includegraphics[width=.9\textwidth]{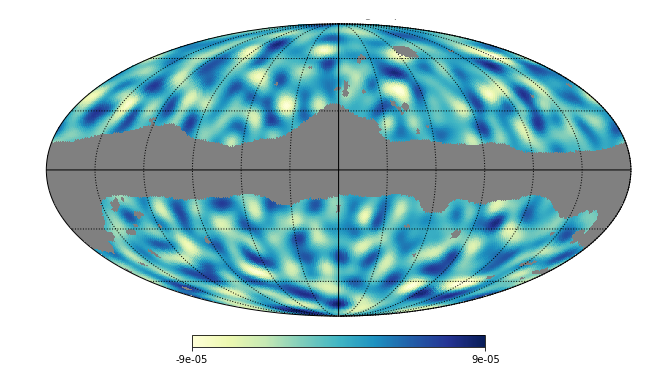}

\caption{\label{filteredmap}  \textit{Planck} lensing map filtered with MH1 at a scale of $5\degree$. The grey region is masked to reduce foreground contamination.
}
\end{figure}

To study the sky at various angular scales, the three Mexican-hat filters described in section~\ref{Mexhat} were applied to the data with
$R={1.0\degree, 1.2\degree, 1.4\degree, \dots, 19.8\degree, 20.0\degree}$, which amounts to 288 ways of filtering the data. Convolving the lensing map with a given filter gives an optimal method for finding features of that specific shape.  One complication is that information from masked regions is smeared out into unmasked pixels. In order to minimize this effect the mask has to be modified. This is achieved by first degrading the mask to $N_\text{side}=128$, then the masked regions are extended by a disk of $0.5R$ around any masked pixel. We should note that we did not mask point sources in this analysis.  
An example of a filtered map is shown in figure~\ref{filteredmap}.  

\subsection{Statistics}
After filtering, we focus on a statistic that counts the area above a given threshold \cite{Planck_13_Isotropy}.  This area is defined as
\begin{subequations}
\label{area}
\begin{align} 
    A^+(R,m)&={\cal N}\{w(R,p)>m\},
    \\
    A^-(R,m)&={\cal N}\{w(R,p)<-m\},
\end{align}
\end{subequations}
which we shall refer to as the `area above' and the `area below', respectively. These are the number of pixels at scale $R$ with wavelet coefficient $w(R,p)$, such that the condition specified in brackets is satisfied. We choose the threshold $m$ to be $4\,\sigma(R)$ (and also test the effect of using $m$ of $3\,\sigma(R)$), where $\sigma(R)$ is the standard deviation of the map filtered at scale $R$. Based on the null hypothesis that the sky should be Gaussian and isotropic on all scales, we can calculate (or assess using simulations) the expected distributions for $A^+(R,m)$ and $A^-(R,m)$.

\section{Results}
\label{Results}
\subsection{Extremal spots in the unmarginalised data}
\label{extremal}

\begin{figure}[htbp!]
    \centering
    \includegraphics[width=\textwidth]{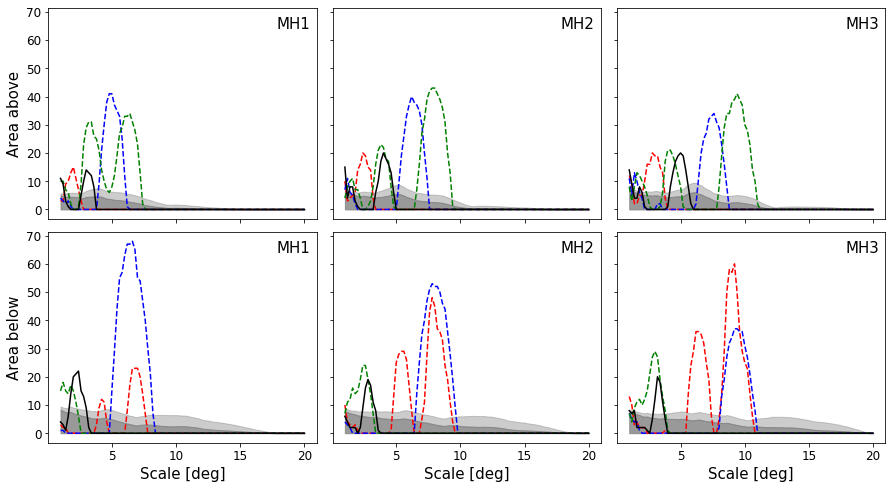}
    \caption{\label{unmargresults1} Results for the $A^+$ and $A^-$ statistics for threshold $m=4\,\sigma(R)$. The labels `MH1', `MH2' and `MH3' stand for the member of the Mexican-hat wavelet family used to filter the data. The dark grey area represents the mean obtained from the set of simulations, whereas the light grey area represents the mean plus 4 times the dispersion in the simulations. The solid black line is the measured data. Scales where the area is zero indicate that no pixels above the threshold are found at that scale. The three coloured dashed lines represent three simulations and, like the data, each simulation exhibits deviations from the mean, well beyond $4\,\sigma$ in both $A^+$ and $A^-$. It is therefore clear that although the data show excursions at some scales, the simulations also show similar excursion, albeit at different scales.}
\end{figure}

\begin{figure}[htbp!]
    \centering
    \includegraphics[width=\textwidth]{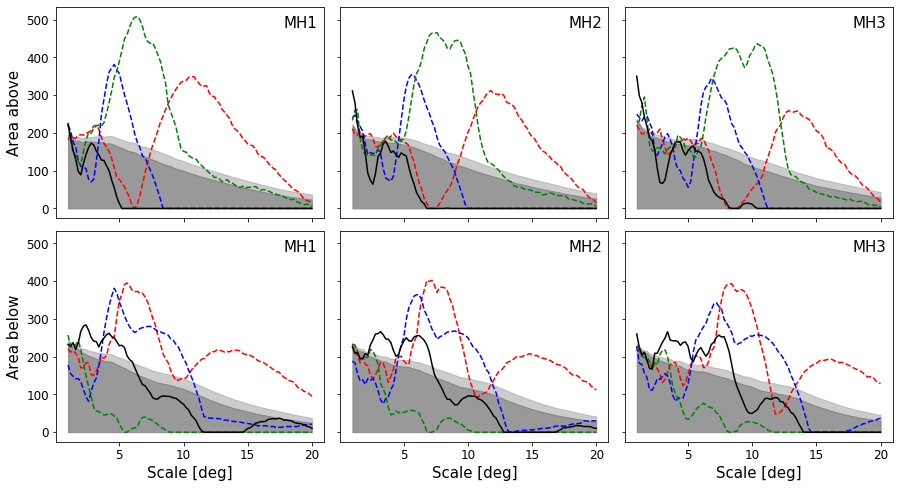}
    \caption{\label{unmargresults2}  Results for the same $A^+$ and $A^-$ analysis as in figure~\ref{unmargresults1}, but for the threshold choice $m=3\,\sigma(R)$. Like in figure~\ref{unmargresults1} the black line (the real data) appears to exceed the expectation in the $A^-$ panels at around $5\degree$, though in this case there are no obvious deviations in $A^+$. Nevertheless, in the same way as in figure~\ref{unmargresults1}, we can easily find simulations (coloured dashed lines) with more extreme spots.}
\end{figure}

We begin our search for extremal spots by comparing the measured data with the results predicted by the simulations. This is achieved by filtering the real data and all the simulated maps, as described in section~\ref{Method}, and calculating $A^+$ and $A^-$ with a threshold of $4\,\sigma(R)$ (as well as $3\,\sigma(R)$) for each filtered map. We then use the 300 provided simulations to find the means and the standard deviations of $A^+$ and $A^-$ for all filters at all scales. To see if there are any statistically significant deviations from the predictions, we compare the measured data to the $4\,\sigma$ confidence interval around the mean predicted by the the simulations. The results of this analysis are presented in figure~\ref{unmargresults1} for $m=4\,\sigma(R)$ and in figure~\ref{unmargresults2} for $m=3\,\sigma(R)$. A point worth stressing is that, since $A^+$ and $A^-$ are by definition non-negative quantities, we are only interested in positive deviations from the predictions, which we represent by plotting the mean $+4\,\sigma$ interval as a light grey band in figures~\ref{unmargresults1} and \ref{unmargresults2}.

\begin{figure}[htbp!]
    \centering
    \includegraphics[width=0.9\textwidth]{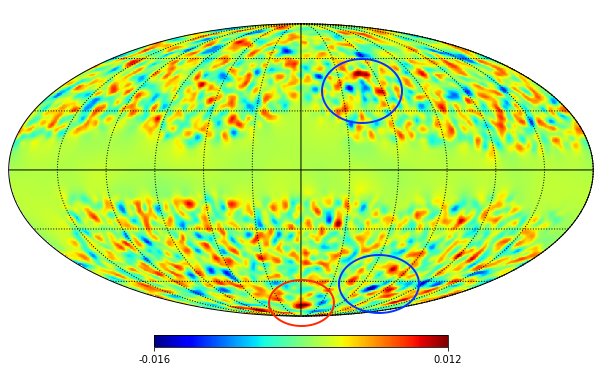}
    \caption{The two most conspicuous cold spots in the lensing map (blue circles) at Galactic coordinates $(l,b)=(325\degree,43\degree)$, $(285\degree, -65\degree$) and the hot spot at $(0\degree, -78\degree)$.  }
    \label{Lensing_cold_spots_mollview}
\end{figure}

Looking at the lensing data (black line) in figures~\ref{unmargresults1} and~\ref{unmargresults2} it might appear that there are anomalies in both $A^+$ and $A^-$ for all filters. In particular, extremal spots in $A^+$ appear to exist at scales between about $2.5\degree$ and $5 \degree$. Similarly, extremal spots in $A^-$ appear to exist at around $2.5 \degree$. By determining which pixels on the sky exceed the $4\,\sigma(R)$ threshold, two specific cold spots were found, located at $(47\degree, 325\degree)$ and $(155\degree, 285\degree)$. Similarly, a hot spot was discovered at $(168\degree, 0\degree)$, as shown in figure~\ref{Lensing_cold_spots_mollview}. However, by looking at simulations (dotted lines), it becomes clear that simulated skies tend to exhibit comparable numbers of hot and cold spots as the real sky.  

In order to test that the results are not coming from the choice of a particular threshold, we also investigated the choice of $3\,\sigma(R)$. Figure~\ref{unmargresults2} shows that with this selection a similar picture emerges as when we used $4\,\sigma(R)$.  Because the threshold is lower than before, a larger number of pixels above the $3\,\sigma(R)$ is found, resulting in more peaks and broader peaks. As was the case for $4\,\sigma(R)$, comparing the data with the mean might appear to suggest that there are unusual cold spots, but similar numbers of spots can be found in the simulations. It may be interesting to note that, in contrast to the $4\,\sigma(R)$ analysis, no hot spots are found here, indicating that an extremal spot found at $4\,\sigma(R)$, does not necessarily stand out at a lower threshold.

\vspace{-0.25cm}

\subsection{Extremal spots in the marginalised data}
\label{marginalised}
From the analysis above one might conclude that there are hot and cold spots in the \textit{Planck} lensing-convergence map. However, this is not the end of the story. Because the MHW filter is not a theoretically predicted shape one could in principle pick other filters to use to search for extrema.  Then in order to ensure that the results are not biased by a posteriori statistics it is crucial to use several filters at different scales and to marginalise over all filters and scales. In other words, one filter and scale may stand out in the data, but we need to check if it stands out more than other filters and scales in the simulated maps. Here we use the data filtered as described in section~\ref{Method}.

To marginalise over the parameter space we first define a quantity that gives the fraction of simulations that are more extreme than the data: 
\begin{equation}
\label{Fm}
     F_m=\frac{{\cal N} [A_{\text{data}}(R',f')>A_{\text{sim}}(R,f)]}{{\cal N}[A_{\text{data}}(R',f' \neq 0 \vee A_{\text{sim}}(R,f)\neq 0]}.
\end{equation}
Here $m$ stands for the threshold described in equation~\ref{area}.  The numerator is the number of scales $(R,R')$ and filters $(f,f')$ at which $A^+$ or $A^-$ for the data is larger than $A^+$ or $A^-$ for a given simulation. This comparison is done between all scales and filters of the data and the simulations for which $A_{\text{data}}(R',f') \neq 0 \vee A_{\text{sim}}(R,f)\neq 0$, hence a scan is performed over the different primed and unprimed labels for scales and filters. The denominator in equation~\ref{Fm} is the number of scales and filters for which the data or the simulation is non-zero.  We calculate $F_m$ for all 300 simulations, using both the $m=4\,\sigma$ and $m=3\,\sigma$ thresholds. 

In order to make it clear how to interpret $F_m$ it is instructive to consider what the distribution of $F_m$ would look like if the data and a set of simulations were drawn from the same distribution. In that case one would expect the frequency of $A_{\text{data}} > A_{\text{sim}}$ to be equal to $A_{\text{data}} < A_{\text{sim}}$ for all scales and filters. Thus, the mean of $F_m$ should be roughly equal to $0.5$, with the simulations forming a symmetric distribution around the mean. To confirm that this marginalisation procedure would pick up on statistically significant signals, we tested it on a fake sky comprised of a simulation with a $6\,\sigma$ signal added to it. As expected $F_{4\,\sigma}$ then has the value of unity for the vast majority of the simulations.

Another way to assess the probability of the existence of extremal spots is to compare the maxima of the area for all simulations with those of the actual data and to calculate the fraction of simulations for which the maximum of the area is bigger for the data than for the simulations. In this way we marginalise by letting each simulation choose its preferred filter scale, i.e. the filter for which the spots seem most striking. 

\begin{figure}[htbp!]
    \centering
    \includegraphics[width=\textwidth]{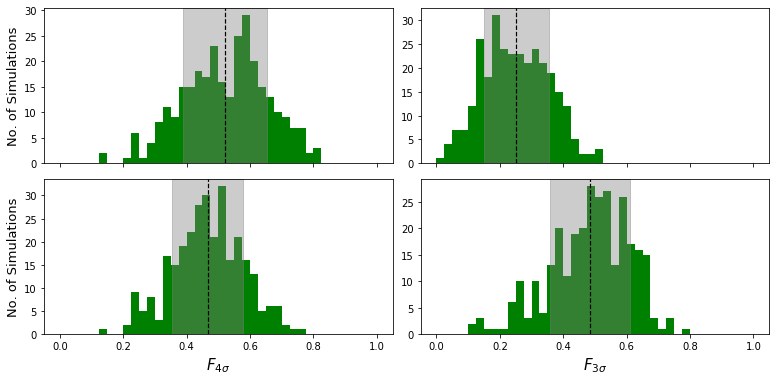}
    \caption{\label{results}Left: Histograms of $F_{4\,\sigma}$ for 300 simulations. The top panel shows the results for $A^+$ and be bottom panel for $A^-$. Both have a mean of around $0.5$, which indicates that there is agreement between the data and the simulations, with no sign of extremal spots that are statistically unusual. Right: Histograms of $F_{3\,\sigma}$ for 300 simulations. The top panel again shows the results for $A^+$ and the bottom panel for $A^-$. For $A^-$ there is no indication of disagreement with simulations, while for $A^+$ there is a roughly $-2.5\,\sigma$ deviation from the $0.5$ expectation. Nonetheless this does not constitute a statistically significant anomaly, given that we searched over both $A^+$ and $A^-$, and could look for both high and low excursions.}
\end{figure}

The main results of the first type of analysis are shown in figure~\ref{results}. Here it should be mentioned that the results are accurate only up to about 1\,\%, because we use 3 different filters at 96 different scales, with the maximum accuracy of $\frac{1}{300}$ limited by the number of available simulations.  The results of the second type of analysis for $m=4\,\sigma$ and $m=3\,\sigma$ are presented in Table~\ref{tab:1}, for each filter separately and for all filters combined.

\begin{table}[htbp!]
\centering
\begin{tabular}{|l|cccc|}
\hline
 &MH1 (\%)&MH2 (\%)&MH3 (\%)& Combined (\%)\\
\hline
$A^+(4\,\sigma)$& 59&66&61&58  \\
$A^-(4\,\sigma)$& 66&56&56&53\\
\hline
$A^+(3\,\sigma)$& 24&74&87&80  \\
$A^-(3\,\sigma)$& 46&37&46&41\\
\hline
\end{tabular}
\caption{\label{tab:1}Fraction of simulations for which the maximum of the area is smaller than for the real data. The top rows are for the $m=4\,\sigma$ threshold and the bottom rows for $m=3\,\sigma$.}
\end{table}

\section{Conclusions}
\label{Conclusion}
We have studied the existence of extremal spots above the threshold $4\,\sigma(R)$ (or $3\,\sigma(R)$) in the \textit{Planck} 2018 lensing data using Mexican-hat wavelet filters. An initial analysis performed by comparing measured data and simulations suggests that apparently extremal spots in the $A^+$ and $A^-$ statistics can be observed at small scales between about $2.5\degree$ and $5\degree$ (and at about $5\degree$ in $A^-$ for the $3\,\sigma$ threshold choice). However, careful marginalisation over all scales and filters leads us to conclude that there is no evidence of statistically significant spots in the lensing convergence maps compared with simulated skies. One might suggest that a larger number of filters could be used, but since the marginalisation procedure in this paper already renders apparent extremal spots statistically insignificant, this is not necessary. A larger number of simulations would be required to improve the accuracy of the results and more sensitive lensing maps would enable searches to be performed over a wider range of scales and shapes.  Additionally if some models made specific predictions for lensing features (in terms of size and shape) then we would not need to marginalise over filters.  It is therefore still possible that there might be some special features in future lensing maps for particular models; however, for now there is no strong evidence for statistically significant spots in lensing data. 

\appendix
\section{Lensing potential}
\label{Lensing Potential}
\begin{figure}[h!]
    \centering
    \includegraphics[width=\textwidth]{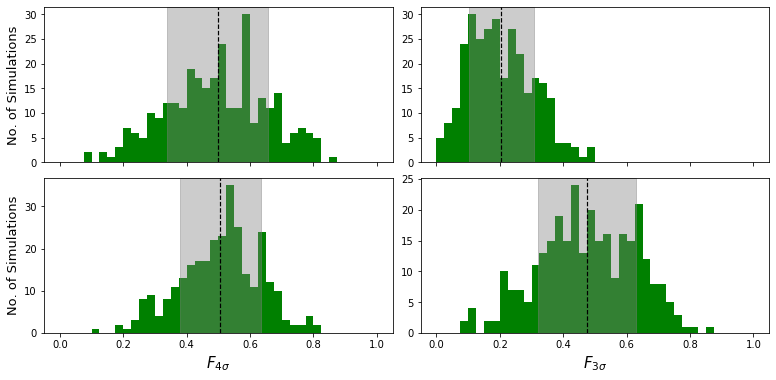}
    \caption{\label{resultspot}Histograms of the area statistics compared with simulations, with the top panels showing results for $A^+$ and the bottom panels for $A^-$.  Left: Histograms of $F_{4\,\sigma}$ for 300 simulations of the lensing potential.   Both have a mean of around $0.5$, which indicates that there is agreement between the data and the simulations, with no sign of extremal spots. Right: Histograms of $F_{3\,\sigma}$ for 300 simulations of the lensing potential.}
\end{figure}

\begin{figure}[htbp!]
    \centering
    \includegraphics[width=\textwidth]{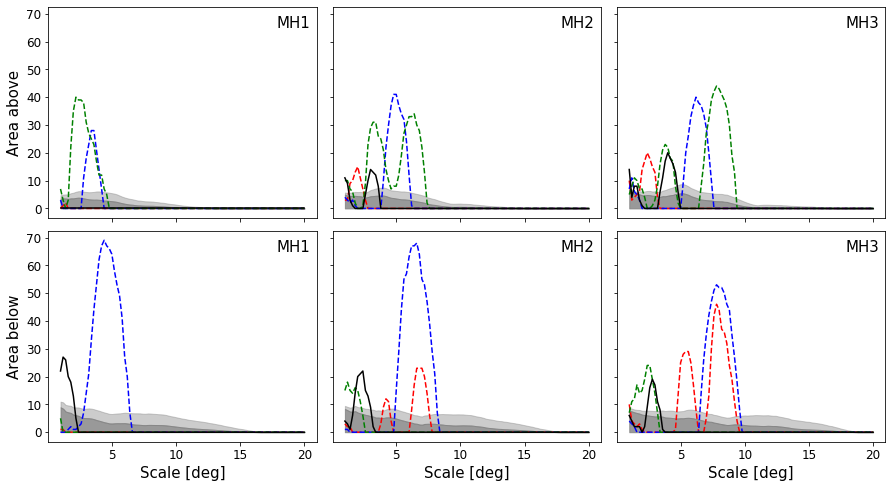}
    \caption{\label{unmargresults1pot}Results for the $A^+$ and $A^-$ analysis of the lensing potential for $m=4\,\sigma(R)$, as in figure~\ref{unmargresults1}. Features at small scales are suppressed as compared to the convergence, as can be seen from equation~\ref{lensingpotential}. For reasons explained in the text, filtering the lensing potential with MH2 is equivalent to filtering the lensing convergence with MH1, and filtering the lensing potential with MH3 is equivalent to filtering the lensing convergence with MH2. Once again the data exhibit deviations larger than $4\,\sigma$, but so do typical simulations.}
\end{figure}

\begin{figure}[htbp!]
    \centering
    \includegraphics[width=\textwidth]{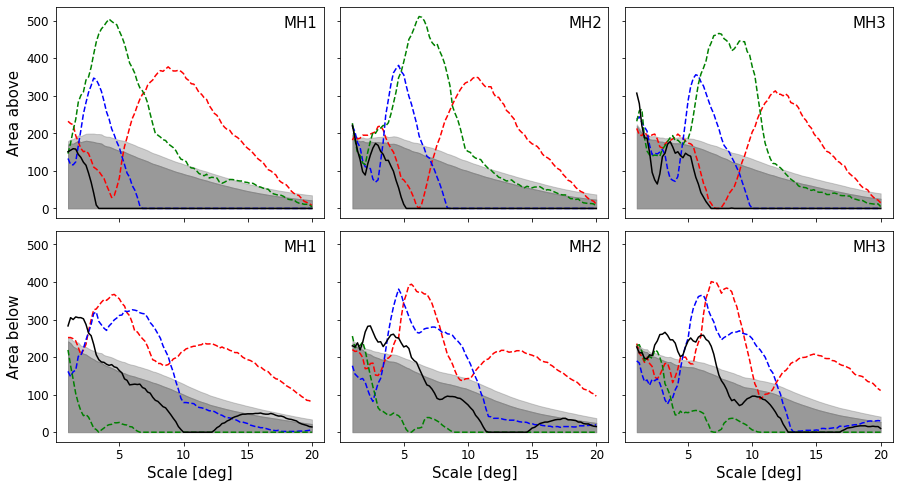}
    \caption{\label{unmargresults2pot}Results for the same $A^+$ and $A^-$ analysis as in figure~\ref{unmargresults1pot} but for $m=3\,\sigma(R)$. Like in figure~\ref{unmargresults1pot} the black line exceeds the $+4\,\sigma$ in the $A^-$ panels around $5\degree$. However, unlike in in figure~\ref{unmargresults1pot}, no excess is found in $A^+$.}
\end{figure}

We shall now briefly review the results of the same analysis repeated on the lensing potential, rather than the lensing convergence. The relation between the lensing potential and the lensing convergence in harmonic space is given by 
\begin{equation}
    \label{lensingpotential}
    \phi_{lm}=\frac{2}{L(L+1)}\kappa_{lm},
\end{equation}
where $\phi_{lm}$ are the spherical harmonic coefficients of the lensing potential and $\kappa_{lm}$ are the spherical harmonic coefficients of the the lensing convergence. Since the three filters used in this paper are related by Laplace operators, as stated in equation~\ref{MHWF}, factors of $(L(L+1)/2$ appear on transformation into harmonic space. This means that
\begin{equation}
    a_{lm}^{\rm MH2}= \frac{L(L+1)}{2} a_{lm}^{\rm MH1}.
\end{equation}
As a consequence, filtering the lensing potential with MH2 is equivalent to filtering the lensing convergence with MH1, and filtering the lensing potential with MH3 is equivalent to filtering the lensing convergence with MH2 (as can be seen in figures~\ref{unmargresults1pot} and \ref{unmargresults2pot}).  This means that for the particular choice of Mexican-hat filters, the potential analysis is not at all independent of the convergence analysis.  The final marginalised results for the potential are presented in figure~\ref{resultspot} and table~\ref{tab:2}. Like for the lensing convergence there is good agreement between data and simulations and no indication for extremal spots. 

\begin{table}[htbp!]
\centering
\begin{tabular}{|l|cccc|}
\hline
 &MH1 (\%)&MH2 (\%)&MH3 (\%)& Combined (\%)\\
\hline
$A^+(4\,\sigma)$& 59&66&61&61  \\
$A^-(4\,\sigma)$& 66&56&56&62\\
\hline
$A^+(3\,\sigma)$& 24&74&87&63  \\
$A^-(3\,\sigma)$& 47&37&47&51\\
\hline
\end{tabular}
\caption{\label{tab:2}Fraction of lensing-potential simulations for which the maximum of the area is bigger for the real data than for the simulations. The top rows are for the $m=4\,\sigma$ threshold and the bottoms rows for $m=3\,\sigma$.}
\end{table}

\clearpage

\acknowledgments
This research was enabled in part by support provided by WestGrid (\url{https://www.westgrid.ca}) Compute Canada (\url{www.computecanada.ca}) and the Natural Sciences and Engineering Research Council of Canada.  It is  based on observations obtained with \textit{Planck} (\url{http://www.esa.int/Planck}), an ESA science mission with instruments and contributions directly funded by ESA Member States, NASA, and Canada.
Some of the results in this paper have been derived using the {\tt HEALPix} package.

\bibliographystyle{JHEP}
\bibliography{Lensing_extrema}

\end{document}